\documentclass[sigconf]{acmart}
\usepackage{algorithm}
\usepackage{siunitx}
\usepackage{url}
\usepackage{multirow}
\usepackage{svg}
\svgpath{{figures/}}

\AtBeginDocument{%
  }

\copyrightyear{2026}
\acmYear{2026}
\setcopyright{cc}
\setcctype{by-nc-nd}
\acmConference[CHI '26]{Proceedings of the 2026 CHI Conference on Human Factors in Computing Systems}{April 13--17, 2026}{Barcelona, Spain}
\acmBooktitle{Proceedings of the 2026 CHI Conference on Human Factors in Computing Systems (CHI '26), April 13--17, 2026, Barcelona, Spain}
\acmPrice{}
\acmDOI{10.1145/3772318.3790685}
\acmISBN{979-8-4007-2278-3/2026/04}

\begin{document}
\title{PuppetChat: Fostering Intimate Communication through Bidirectional Actions and Micronarratives}

\author{Emma Jiren Wang}
\email{jiren@vt.edu}
\orcid{0009-0003-4413-9944}
\affiliation{%
  \institution{Virginia Tech}
  \city{Blacksburg}
  \state{Virginia}
  \country{USA}
}

\author{Siying Hu}
\email{siyinghu@cityu.edu.hk}
\affiliation{%
  \department{Department of Computer Science}
  \institution{City University of Hong Kong}
  \city{Hong Kong}
  \country{SAR}
}

\author{Zhicong Lu}
\email{zlu6@gmu.edu}
\affiliation{%
  \department{Department of Computer Science}
  \institution{George Mason University}
  \city{Fairfax}
  \state{Virginia}
  \country{USA}
}
\renewcommand{\shortauthors}{Wang et al.}

\newcommand{\sy}[1]{\textcolor{cyan}{#1}}

\begin{abstract}
As a primary channel for sustaining modern intimate relationships, instant messaging facilitates frequent connection across distances. However, today's tools often dilute care; they favor single tap reactions and vague emojis that do not support two way action responses, do not preserve the feeling that the exchange keeps going without breaking, and are weakly tied to who we are and what we share.
To address this challenge, we present PuppetChat, a dyadic messaging prototype that restores this expressive depth through embodied interaction. PuppetChat uses a reciprocity aware recommender to encourage responsive actions and generates personalized micronarratives from user stories to ground interactions in personal history. Our 10-day field study with 11 dyads of close partners or friends revealed that this approach enhanced social presence, supported more expressive self disclosure, and sustained continuity and shared memories.
\end{abstract}




\begin{teaserfigure}
  \includegraphics[width=\textwidth]{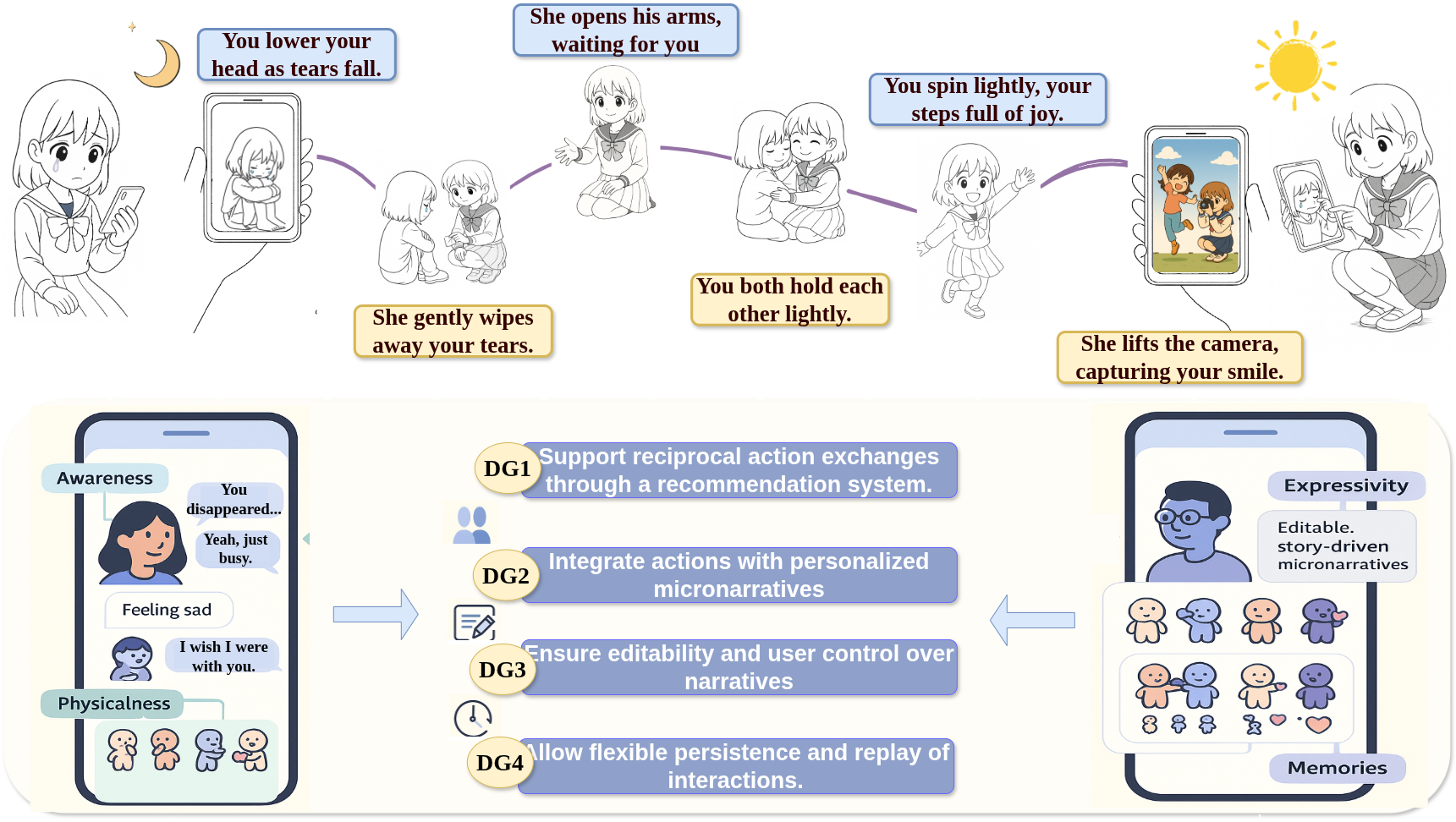}
  \caption{\textbf{Conceptual vision and interface implementation of the expressive messaging system. (Top) A conceptual story sequence illustrating the types of emotional support and reciprocal physical interactions the system aims to facilitate between users. (Bottom) Interface mockups demonstrating the actual translation of these concepts into mobile UI features. The center lists four key Design Goals (DG1–DG4). Arrows indicate how specific UI components, such as the recommended avatar actions under "Physicalness" (implementing DG1 \& DG2) and the saveable interaction history under "Memories" (implementing DG4), directly address these design goals.}}
  \Description{Conceptual vision and interface implementation of the expressive messaging system. (Top) A conceptual story sequence illustrating the types of emotional support and reciprocal physical interactions the system aims to facilitate between users. (Bottom) Interface mockups demonstrating the actual translation of these concepts into mobile UI features. The center lists four key Design Goals (DG1–DG4). Arrows indicate how specific UI components—such as the recommended avatar actions under "Physicalness" (implementing DG1 and DG2) and the saveable interaction history under "Memories" (implementing DG4)—directly address these design goals.
}
  \label{fig:teaser}
\end{teaserfigure}


\maketitle
\section{Introduction}
Instant messaging (IM) has become a primary medium for sustaining relationships across distance, yet in this setting intimate expression is often diluted by the logic of convenience \cite{instant01, verheijen2013effects, tyson2005instant}: when people truly need to be understood, comforted, or “do something together,” communication is compressed into a handful of icons or a block of text that takes significant effort to compose, either too light and easy to misread or too heavy and hard to write. Conversations frequently resolve into non-invitational and terminal feedback \cite{coherence01, schwarz2011moved}, such as long pressing a message to attach a thumbs-up reaction, which neither conveys care promptly nor offers a way to link the present moment with prior shared experience \cite{challenge01, groffman2010restarting}. 
Prior work shows that feeling “we are here together” is crucial for maintaining relational closeness, emotional support, and shared meaning in mediated communication \cite{socialpresence01}. Without this sense of presence, communication may appear efficient yet still fail to nurture connection.

To enhance relatedness under spatial separation, prior work highlights a set of complementary strategies, namely \textit{Awareness, Expressivity, Physicalness, Joint Action, and Memories} \cite{allYouNeedisLove, falbo1980power, overall2009regulating}. Taken together, they argue that intimate communication should not prioritize convenience alone; it should gently make partners feel "seen," enable emotional expression that is personalized yet leaves room for interpretation, restage care through appropriate embodied cues \cite{embodied01, wmbodied02}, transform replies from endpoints into invitations to the next move to \textit{sustain continuity} \cite{allYouNeedisLove, fampracmp099}, and accumulate moments into a revisitable shared timeline. Guided by this lens, we take these elements as design levers to reconfigure everyday IM for intimacy.

We present PuppetChat, a dyadic messaging prototype that operationalizes these elements in everyday chat. 
To support \textit{awareness}, PuppetChat offers low friction action-plus-micronarrative exchanges, where an action is a short expressive puppet animation (e.g., hug, wipe-tears) and a micronarrative is a brief caption style text that contextualizes the action, allowing a partner's momentary state to surface without demanding attention.
For \textit{expressivity}, it generates editable, story driven micronarratives anchored in user provided personal stories, allowing meanings to be tailored beyond generic emojis. To approximate \textit{physicalness}, PuppetChat uses embodied puppet avatars capable of performing 42 actions (e.g., hug, wipe-tears) that restage care rather than merely label it. 
To support joint action, the system uses an LLM based recommender that suggests actions that naturally complement the partner's action (e.g., when one partner selects cry, the system may suggest wipe-tears). This turns individual actions into opportunities to co-perform.
Finally, to cultivate \textit{memories}, actions and micronarratives are preserved in the thread as lightweight traces that can be revisited, helping interactions accumulate into a shared timeline.

To evaluate how PuppetChat supports intimate expression, joint action, and relational presence in everyday messaging,across a 10-day field study with 11 dyads of close partners (N=22), we found that reciprocal actions reliably increased social presence and a felt sense of co-performance even in a non-immersive, text based interface. Lightweight and responsive actions paired with micronarratives turned replies into sequences rather than one-off reactions, maintaining the flow of interaction instead of breaking it into disconnected turns.
Editable and story driven micronarratives afforded precise yet face saving self disclosure and seeded playful co-construction of meaning (e.g., inside jokes and private riffs). Over time, the accumulation of actions and captions created revisitable traces that supported reflection on shared moments. Taken together, these effects lowered the friction of nuanced expression while strengthening reciprocity and continuity, which ultimately enhancing partners' sense of relatedness in intimate relationships.

Our research thus makes three primary contributions to the HCI community:

\begin{itemize}
\item \textbf{A novel interface paradigm for intimate communication that integrates embodied action with personalization micronarratives.} We present PuppetChat, a system demonstrating how lightweight bidirectional avatar actions and AI generated micronarratives can support
more reciprocal and narratively grounded dyadic chat.
\item \textbf{Empirical insights into how reciprocal and embodied interactions foster connection in mediated contexts.} Our field study identifies mechanisms by which users employ responsive actions for "co-performance" and editable narratives for playful "co-construction" of meaning, thereby strengthening relatedness in intimate relationships.
\item \textbf{Design implications.} We outline concrete directions for technologies that move beyond one directional reactions to foster deeper narrative depth and interactional continuity in digital relationships.
\end{itemize}

\section{Related Work}
This work reviews prior studies in three key areas: (1) \textit{awareness and social presence}, which investigate how peripheral and ambient cues foster a sense of "being seen" and "being together" across distance; (2) \textit{physicalness and joint action}, which examine the use of physiological proxies and re-enacted gestures or rituals to recreate bodily co-presence within instant messaging contexts; and (3) \textit{expressivity and micronarrative}, which explore enriched communicative forms ranging from minimal affective signals to dyad specific symbolic languages and short, revisable narratives that bridge lightweight cues and extended messages.

\subsection{Awareness and Social Presence}
To evoke a sense of "being seen" and "being together" in intimate relationships, a large body of work employs peripheral, low intrusion cues that permeate everyday settings: rather than relying on explicit conversation, signals are embedded in spaces and objects so that information flows continuously in the background \cite{oh2018systematic, lowenthal2010social}. Interfaces and artifacts in this line prioritize being felt at the periphery over commanding center stage attention \cite{shils1961centre}, using light, implicit cues to cultivate social presence. For example, merely indicating that "the other is there," diffusely mapping everyday activity rhythms, or conveying affect through environmental/physiological hints \cite{sproull1986reducing, cues01}. Unlike message centric approaches, the value here lies in continuity without interruption: the partner's existence and routines are sensed in a subtle, always on manner that folds into daily practice, instead of launching a new explicit exchange.

Empirical studies show that such "small yet continuous" cues can heighten state awareness and connectedness, and help counter over idealization in long distance contexts by supporting a more realistic mutual understanding \cite{allYouNeedisLove, socialpresence01, socialpresence02}. Turning awareness into a positive experience, however, depends on matching intensity, frequency, and audience of disclosure to relationship stage and situation: cohabiting versus long distance, work versus public settings, all differ in tolerance for sound, ambience, and rhythm cues \cite{RIORDAN20101806, makinglove}. Consequently, prior work often stresses interpretable ambiguity and boundary control \cite{fattorini1968boundary}: leaving room for receiver interpretation to reduce feelings of surveillance, while allowing the sender to adjust visibility (to whom, when, at what granularity). At the same time, limitations emerge. Many awareness signals remain one way indicators weakly coupled to conversational turn taking \cite{allYouNeedisLove, affectivecommunication}. They increase being seen but do not necessarily invite the next move, and they rarely connect systematically to relationship narratives or shared memory \cite{alea2007ll}.

Taken together, peripheral awareness cues succeed at sustaining "being seen," yet they often stop at one directional indication, insufficiently integrated with IM's turn taking mechanics and thus poor at inviting follow up interaction. They also face a trade off between attentional burden and intrusiveness, where signals that are too subtle are overlooked \cite{swallow2013attentional}; too explicit and they disrupt the moment, and their contextual fit and disclosure granularity are frequently not tunable across relationship stages and public or private settings. In response, we embed awareness cues within the IM dialogue flow via low friction, peripherally perceivable action prompts (which can be sent action only for weak disclosure and minimal disruption) and use complementary response mechanisms to move from merely "being seen" to being received, encouraging a next turn with symmetric or complementary actions. This reframes presence from a static indicator into respondable co-presence, while offering controllable intensity and visibility (e.g., ephemeral status versus more persistent options) so that salience and interactivity improve without sacrificing privacy boundaries. In this way, awareness becomes more than a faint background signal: it lightly activates reciprocity and a felt sense of togetherness, thereby more effectively supporting social presence.

\subsection{Physicalness and Joint Action}

When partners are apart, the most palpable loss is the sense of bodily co-presence \cite{co-presence, hanks2013counterparts}. Prior work responds along two routes: one uses \textit{physiological proxies} (e.g., heart rhythm, skin warmth, rhythmic vibration) to evoke the bodily side effects of proximity \cite{morgan1943physiological}; the other \textit{re-enacts gestures} (e.g., holding, stroking, hugging) through wearables or actuated objects to stage experiences of being held or cared for \cite{wearable01, wearable02, wearable03, wearable04}. In parallel, research on \textit{joint action} seeks to re-establish a rhythm of doing together \cite{jointaction01}, either by inventing new dyadic rituals \cite{joinaction02} (e.g., small coordinated activities across bedrooms or remote mini games) or by piggybacking on mundane routines \cite{jointaction03} (e.g., synchronized sipping). The two lines share a goal: physicalness brings back the felt texture of care, while joint action recreates a shared tempo \cite{allYouNeedisLove}.

Touch is a direct conduit for empathy and care, yet mediation introduces stubborn frictions \cite{hertenstein2009communication, touch01}. \textit{Immediacy} is difficult: embraces and hand holds rely on prompt back and forth, whereas technical delays sap reciprocity \cite{andersen1979measurement, vasseleu1999touch}. \textit{Context fit} is fragile: what feels tender in private can appear conspicuous or awkward in public, pushing users to abandon visible devices \cite{visible01}; many systems therefore pivot from one-to-one haptics to \textit{symbolic or poetic} evocations of care \cite{hapic01, hapic02}. On the joint-action side, the target is \textit{behavioral interdependence} \cite{wageman2014meaning, rusbult2008we,interindependence, interdepence02}, but many implementations settle into parallel play or light synchronization that couples weakly to IM turn taking and rarely \textit{invites a next move} \cite{interpersonal03}. Selecting activities that both trigger mutual influence and embed smoothly in everyday life across settings remains a central design tension.

In sum, mediated touch can evoke closeness but struggles with synchronous reciprocity and situational appropriateness; joint-action systems foster togetherness yet often depend on artificial rituals or weak interdependence. What is missing in IM contexts is a \textit{low friction, turn embedded} means to \textit{restage care as action} and to \textit{transform replies from endpoints into invitations}, without demanding strict synchrony or heavy hardware. We address this by integrating \textit{embodied metaphors of care} directly into reply structure and leveraging \textit{complementary actions} (e.g., distressed move \textrightarrow{} soothing response) so that physicalness is enacted within the chat's turn system and joint action emerges as a reciprocal rhythm. This keeps salience and intrusiveness tunable while foregrounding behavioral interdependence in everyday messaging.

\subsection{Expressivity and Micronarrative}
Expressive communication in intimate contexts goes beyond merely "saying how one feels"; it also uses encoded or enriched forms that make feelings visible, answerable, and even playful \cite{express01, gross1998mapping, hubler2011communication, staahl2005foundation}. Prior systems span two broad lines: one centers on \textit{minimal on--off signals} \cite{signals01}(e.g., a press, a squeeze mapped to light or vibration) to create ultra low friction affective touchpoints in both synchronous and asynchronous use \cite{touchpoints01}; the other invites partners to co-create a \textit{dyad specific symbolic language} \cite{dearboard, customization}, composing messages from shared symbols \cite{fay2018create, fowler1987shared, simpson1990influence} whose meanings only the two of them understand. Both lines rely on \textit{interpretable ambiguity}: the same cue can take on different meanings by context, preserving discretion in public while sustaining a sense of "us" \cite{baxter1986turning, romantic01, romantic02}.

Relationship research underscores that \textit{emotional expression is necessary} for maintenance \cite{emtional01, emotional02, furman1997adolescent, keltner2019emotional}; suppressing it incurs cognitive costs \cite{keltner2019emotional}, and in long distance settings more frequent affective writing is linked to stability \cite{longdistance01}. Intimate exchange also carries a strong \textit{reciprocity expectation} \cite{burgoon1993effects, rosenthal1967unintended}: expressions call for timely replies, so tools must embed into daily rhythms \cite{li2018daily}, lowering the threshold to speak and making it easy to be "caught." \cite{nisan1993communication} Yet tensions persist. Minimal cues often collapse into \textit{terminal feedback} (a tap that ends the exchange), symbol systems can \textit{miscalibrate tone} or be misread, and people differ in expressing and recognizing emotion; gender, culture \cite{genderexpression, genderandculture}, and the handling of \textit{negative affect} (argument, repair) are frequently under addressed \cite{jones2011relational}.

Between low friction but semantically thin signals and nuanced but effortful long text, IM lacks a \textit{context coupled, editable middle layer}. Short, revisable \textit{micronarratives} \cite{digra669, micronarrative2024} address this gap: a line or two tethered lightly to the moment (and optionally to an action) that \textit{clarifies tone, marks intent, and preserves face} while \textit{inviting a next move}. Unlike fixed symbols, micronarratives emphasize \textit{rewritability and fit to context} \cite{digra669}; unlike full messages, they remain \textit{lightweight and improvisable}. In everyday IM, such units can shift expression from a one way display to a \textit{turn worthy} contribution, providing fine grained semantic footholds for continued, reciprocal interaction.

\section{Formative Study}
We adopted an iterative design process. As a first step, we conducted a small scale formative study using an early demo prototype of PuppetChat to explore users' initial reactions, which informed the design of the more complete prototype developed afterward.
\subsection{Participants} 
We recruited 5 participants (3 women and 2 men) for this study. All participants were frequent instant messaging users who regularly engaged in dyadic chats to sustain their daily social and relational communication. None of them had prior exposure to our system. Participants reported using a range of IM platforms in their everyday life, and were familiar with common expressive resources such as emojis, stickers, and GIFs.


\subsection{Study Procedure}
We used a prototype demonstration \cite{bodker2020design} method to gather early reactions to PuppetChat's core interaction concepts. Each 30–40 minute session began with an informed consent and a brief introduction to the study objectives. Participants were then shown an early demo prototype designed to present only the minimal idea of using animated actions in chat.

The prototype used an abstract purple avatar to avoid implying gender or identity and keep attention on the interaction concept. We included twelve actions that provided a small but expressive set of relational gestures, selected through early sketching and informed by prior work on expressive messaging\cite{instant01, hubler2011communication, staahl2005foundation}.
Each action GIF displayed a \textit{short action description} on hover, enabling participants to confirm the intended meaning of the action during exploration. The interface allowed participants to enter a short text message, after which the system used an LLM to analyze the input and recommend four candidate actions (Fig.~\ref{fig:workflow} A1). Selecting one of these options triggered the corresponding GIF animation. For comparison purposes, we also showed two sample animations (throw-heart and carry-heart) from the human-like girl avatar used in the finalized system and from a robot avatar, although these were presented only briefly at the end of the session rather than as part of the interactive prototype. The demo did not include \textit{personalized micronarratives, personal stories, a replay function}, or \textit{complementary action recommendations} based on a partner's actions, allowing us to probe user expectations without introducing predetermined design directions.

Participants were asked to (1) enter short messages, (2) review the recommended actions, and (3) trigger action animations using the purple avatar. After exploring these interactions, a semi-structured interview invited reflections on their expectations, preferences, and everyday IM practices. During the interview, participants were also shown the two additional avatar examples and asked for brief feedback. They further provided examples of how they currently use emojis, stickers, and GIFs, and discussed how animated actions might extend or disrupt these existing practices.

\subsection{Analysis}
Interview notes and transcripts were analyzed using a lightweight thematic analysis approach\cite{braun2019reflecting}. Two authors independently reviewed the session notes, generated codes based on what appeared meaningful in the data, and refined these through discussion until agreement was reached. The aim of the analysis—given the exploratory nature of the formative study—was to surface notable user expectations and potential design opportunities.

\subsection{Findings}

Our findings partly aligned with expectations grounded in prior literature. For example, users' desire for reciprocal exchanges (F1) reflects well established observations about turn taking and mutual responsiveness in mediated communication \cite{allYouNeedisLove, ling2008newtech}. Similarly, the value of personalization and contextual cues (F2–F3) echoes prior work on intimacy and relational maintenance in messaging systems \cite{customization}.
At the same time, the study surfaced several unanticipated needs, such as the desire for flexible persistence (F4), which were not explicitly highlighted in earlier research. These insights informed the design of the full prototype and extended the system beyond what prior work alone would have predicted.

\textbf{F1: Users sought reciprocal exchanges and clearer guidance for responses} 

Participants emphasized that expressive actions should not be isolated, one-off tokens like emojis or stickers. Instead, they wanted actions to invite a partner's response, creating a closed loop of interaction. Several participants described frustration when an action such as a "hug" felt incomplete without a corresponding reply, noting that \textit{"if the other person doesn't respond, it feels unfinished."} To address this, they suggested that the system could highlight when a response was received or recommend complementary actions to make reciprocity easier—for example, pairing "throw-heart" with "catch-heart" or suggesting playful alternatives such as "split-heart" or "throw back the heart." These ideas point to the importance of embedding reciprocal structures into action design, as well as supporting intelligent recommendations to encourage mutual responsiveness.

\noindent \textbf{F2: Users wanted actions to be grounded in personal narratives} 

Participants felt that actions alone risked being too generic or ambiguous, and that pairing them with short narratives could add clarity and intimacy. As P3 noted, \textit{"If I just send a pat on the shoulder, it feels flat. But if I can add a line like 'just like that day after your exam,' it becomes ours."} Similarly, P1 remarked, \textit{"Sometimes an action might be misread---like is this playful or serious? A few words can make the meaning clear."} Others emphasized the potential of narratives to embed personal identity and relationship specific themes. For instance, P2 explained, \textit{"I'd love to add our inside jokes or little phrases we always use---it makes the chat feel more like us." }Overall, participants saw the combination of action and micronarrative as a way to transform fleeting gestures into more meaningful and accumulative exchanges.

\noindent \textbf{F3: Users desired flexibility to edit and control generated narratives} 

Because relational communication is highly personal, participants expressed a strong desire for flexibility in shaping both actions and narratives. They wanted the ability to \textit{edit, refine, or override AI generated text}, as well as to adjust persona cues that might influence how a message is framed. As P2 explained, \textit{"Sometimes the AI sounds too formal---I just want to tweak it a little so it feels like my own voice."} Similarly, P4 remarked, \textit{"I don't want to be forced into one option. It's important I can decline a suggestion or swap it for something that feels right for us."} These preferences highlight that users want AI to augment their expressiveness, but not to replace their agency.

\noindent \textbf{F4: Users expressed mixed preferences for ephemeral vs. persistent interactions} 

Participants expressed a strong desire for flexibility in how expressive interactions are shared and retained. They wanted the option to send an action without any accompanying narrative, treating it as a lightweight and ephemeral gesture. At the same time, they emphasized the value of being able to choose whether an interaction should vanish quickly or persist as a lasting record. As P2 explained, \textit{"Sometimes I just want it to disappear like a quick reaction, but other times I want it to stay so we can come back to it."} When interactions were made persistent, participants hoped they could also revisit the experience by replaying the associated action. P4 noted, \textit{"If it's saved, I'd like to click on it and see the moment again, not just read the text."} These preferences highlight users' desire to balance spontaneity with continuity, allowing them to decide when an exchange should be transient and when it should become part of their shared history.

\subsection{Design Goals}

Based on the findings from our formative study (F1–F4) and insights from prior literature on expressive messaging, social presence, and relational communication \cite{allYouNeedisLove, cassell2000, birnholtz2014} 
we distilled the following design goals to guide the development of PuppetChat. 
Each design goal directly addresses one of the key findings.  

\noindent \textbf{DG1. Support reciprocal action exchanges through a recommendation system}.  
The system should provide interactive actions that invite reciprocal responses, supported by a recommendation mechanism. Based on a scoring process, the system should suggest four candidate actions tailored to the conversational context (e.g., recommending "catch-heart" after "throw-heart"), enabling playful turn taking and mutual responsiveness in dyadic chat. (F1)

\medskip
\noindent \textbf{DG2. Generate micronarratives grounded in users' personal stories.}
The system should create short, customizable narrative cues that draw on users' personal story inputs and the selected action, allowing each message to carry contextually rich, relationship specific meaning. (F2, F3)

\medskip
\noindent \textbf{DG3. Ensure editability and user control over narratives.}
The system should allow users to edit, override, or reject AI generated narratives and recommendations, ensuring personalization while preserving user agency. (F3)

\medskip
\noindent \textbf{DG4. Allow flexible persistence and replay of interactions.} 
The system should give users control over whether interactions remain ephemeral or are retained as revisitable records. When retained, interactions should also support an action replay function, allowing users to re-experience the associated gesture. This flexibility balances spontaneity with the creation of lasting and memorable exchanges. (F4)


\section{PuppetChat System}
To address the 4 design goals identified in our formative study, 
we designed and implemented \textit{PuppetChat}, a real time one to one conversational platform that augments instant messaging with reciprocal action exchanges and personalized micronarratives.  
\textit{PuppetChat} augments instant messaging with reciprocal action exchanges paired with customizable \textit{micronarratives}---short textual annotations that describe and contextualize each action in a personal way. One action–micronarrative pair provokes a reply, which in turn prompts another, creating a continuous chain of exchanges rather than isolated reactions. 

\begin{figure*}[htbp] 
    \centering
    \includegraphics[width=\textwidth]{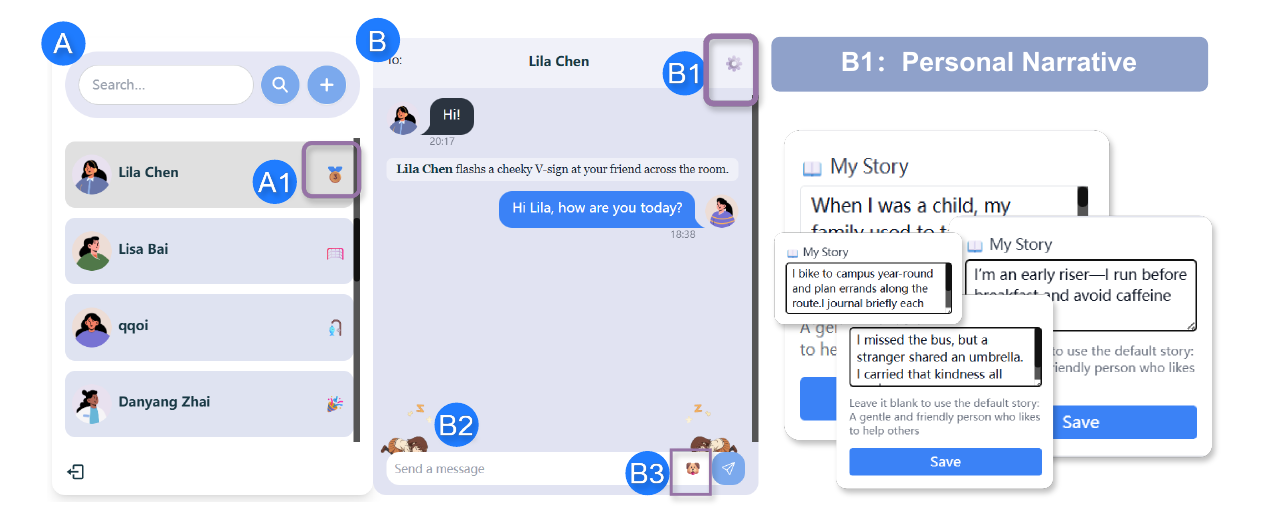}
    \caption{\textbf{The main interface of \textit{PuppetChat}. (A) The contact panel supports user management. (A1) Relationship management through personalized icons. (B) The conversation panel for dyadic exchange. (B1) Personal narrative input for guiding and editing AI generated micronarratives with persona cues. (B2) The interactive puppet area, where puppets animate actions or remain idle in a resting pose (left: partner; right: self). (B3) Action recommendation button that suggests reciprocal actions based on conversational context.}}
    \Description{The main interface of PuppetChat. (A) The contact panel supports user management. (A1) Relationship management through personalized icons. (B) The conversation panel for dyadic exchange. (B1) Personal narrative input for guiding and editing AI-generated micronarratives with persona cues. (B2) The interactive puppet area, where puppets animate actions or remain idle in a resting pose (left: partner; right: self). (B3) Action recommendation button that suggests reciprocal actions based on conversational context.}
    \label{fig:main_page}
\end{figure*}

The primary user interface of \textit{PuppetChat}, as depicted in Figure~\ref{fig:main_page}, 
serves as the central workspace for composing, sending, and responding to action–narrative messages. 
As shown in Figure~\ref{fig:main_page}, the main interface of \textit{PuppetChat} consists of two primary components: the contact panel (Fig.\ref{fig:main_page} A) and the conversation panel (Fig.\ref{fig:main_page} B). The contact panel (Fig.\ref{fig:main_page} A) enables users to add or search for contacts and manage login or presence status; within it, the relationship management module (Fig.\ref{fig:main_page} A1) allows users to assign different icons to organize and differentiate among friends, family, partners, or varying degrees of closeness. The conversation panel (Fig.\ref{fig:main_page} B) integrates multiple functions for dyadic interaction: the personal narrative input (Fig.\ref{fig:main_page} B1) allows users to enter individualized cues and persona settings, which guide AI in generating and refining micronarratives; the interactive puppet area (Fig.\ref{fig:main_page} B2) visualizes actions, where puppets remain in a resting pose when inactive and animate once an action is triggered, with the partner's puppet displayed on the left and the user's on the right; and the action recommendation button (Fig.\ref{fig:main_page} B3) provides suggested actions based on conversational context, supporting reciprocal exchanges.

In this section, we describe the system design of \textit{PuppetChat} in four parts. 
We begin by discussing how the action library was curated and annotated then outline key features of \textit{PuppetChat}. 
Next, we conclude with the technical implementation that enables real time multimodal interaction.

\subsection{System Design}
This section outlines the design rationale and implementation choices behind PuppetChat’s avatar and action system.
\subsubsection{Avatar Design}
The finalized prototype uses a human-like girl avatar as its primary expressive character. This choice was informed by three considerations.

(1) Our formative study showed that participants, regardless of gender, responded more positively to human-like characters than to robot or abstract avatars, and none of the male participants expressed discomfort with the girl avatar. 
(2) Prior research on 2D expressive assets, including stickers, emojis, and animated GIFs, indicates that user adoption is primarily driven by emotional clarity, stylistic expressiveness, and perceived enjoyment, rather than the specific gender of the character \cite{sawmong2022role}.
In text based communication, these assets function as paralinguistic cues or pragmatic markers to modify tone, rather than as embodied self representations or strict identity claims \cite{tang2019emoticon, aumuller2024rethinking}. Consequently, the imperative for gender matching, while critical in embodied VR contexts to maintain presence, is significantly diminished in messaging systems. Evidence suggests that in these 2D environments, users prioritize functional utility and the ability to convey abstract concepts over demographic alignment, often preferring neutral or abstract representations when given the choice.
(3) A further practical factor was technical feasibility. The system's action library includes 42 expressive actions that require precise motion alignment, timing, and consistency. Generating and maintaining separate animation sets for multiple avatars would significantly increase production workload, model size, and integration complexity. Given our focus on evaluating core interaction mechanisms rather than character customization, we opted to standardize on a single human-like avatar to ensure consistency across actions and reduce implementation overhead. This decision mirrors common practice in animated sticker systems, which typically adopt a single stylized character for production efficiency.
For these reasons, we adopted a single stylized human-like character, while acknowledging that offering additional avatar options is a valuable direction for future work.

\subsubsection{Designing the Action Set}

The action library was constructed through a multi stage curation and annotation process. We began by collecting 215 animated GIFs featuring clearly identifiable protagonists and distinct actions, sourced from popular meme repositories as well as two dedicated GIF sharing websites\footnote{\url{https://giphy.com/}}\footnote{\url{https://tenor.com/}}. Each action was first categorized along two orthogonal dimensions: \textit{emotional valence} (neutral, positive, negative) and \textit{interaction role} (self oriented vs.\ responsive). This structure enabled the definition of potential \textit{ReactionCandidate} pairings (e.g., "throw-heart" can be paired with "carry-heart" as a complementary response, while "cry" can be paired with "wipe others' tears").  

To ground these categories in everyday practice, we further collected over 20 conversation logs comprising nearly 100 instances of animated GIFs, emojis, and emoticons. We performed an inductive qualitative analysis on these instances, where two researchers independently coded the communicative intent of each artifact based on the surrounding textual discourse. From this process, we curated and refined a library of 42 representative actions, balancing expressive range with practical usability.  

Each action in our library is represented as a structured object with attributes such as \textit{name}, \textit{description}, \textit{keywords}, \textit{emotion}, \textit{interaction role}, \textit{embedding array}, and \textit{reactionCandidates}; details are summarized in Table~\ref{tab:action-attributes}.

\begin{table*}[t]
\centering
\small
\caption{\textbf{Attributes of each action in the library.}}
\Description{This table lists the attributes defined for each action in our system’s library. Each action is represented as a structured object that includes: the Name, serving as the label of the action; the Description, a concise textual summary of the action’s content and expressive intent; Keywords, terms that capture typical triggers or contexts; the associated Emotion category; the Interaction role, situating the action within dyadic interaction; an Embedding array, a high-dimensional vector that enables semantic fuzzy matching; and ReactionCandidates, which link to complementary or reciprocal actions.}
\label{tab:action-attributes}
\begin{tabular}{p{3cm}p{10cm}}
\toprule
\textbf{Attribute} & \textbf{Description} \\
\midrule
Name & The label of the action. \\
Description & A concise textual summary of the action's content and expressive intent. \\
Keywords & A small set of terms that capture typical triggers or contexts. \\
Emotion & The associated affective category. \\
Interaction role & The situational function of the action within dyadic interaction. \\
Embedding array & A high dimensional vector enabling semantic fuzzy matching. \\
ReactionCandidates & Links to complementary or reciprocal actions. \\
\bottomrule
\end{tabular}
\end{table*}

\subsection{Key Features}

\begin{figure*}[htbp] 
    \centering
    \includegraphics[width=0.8\textwidth]{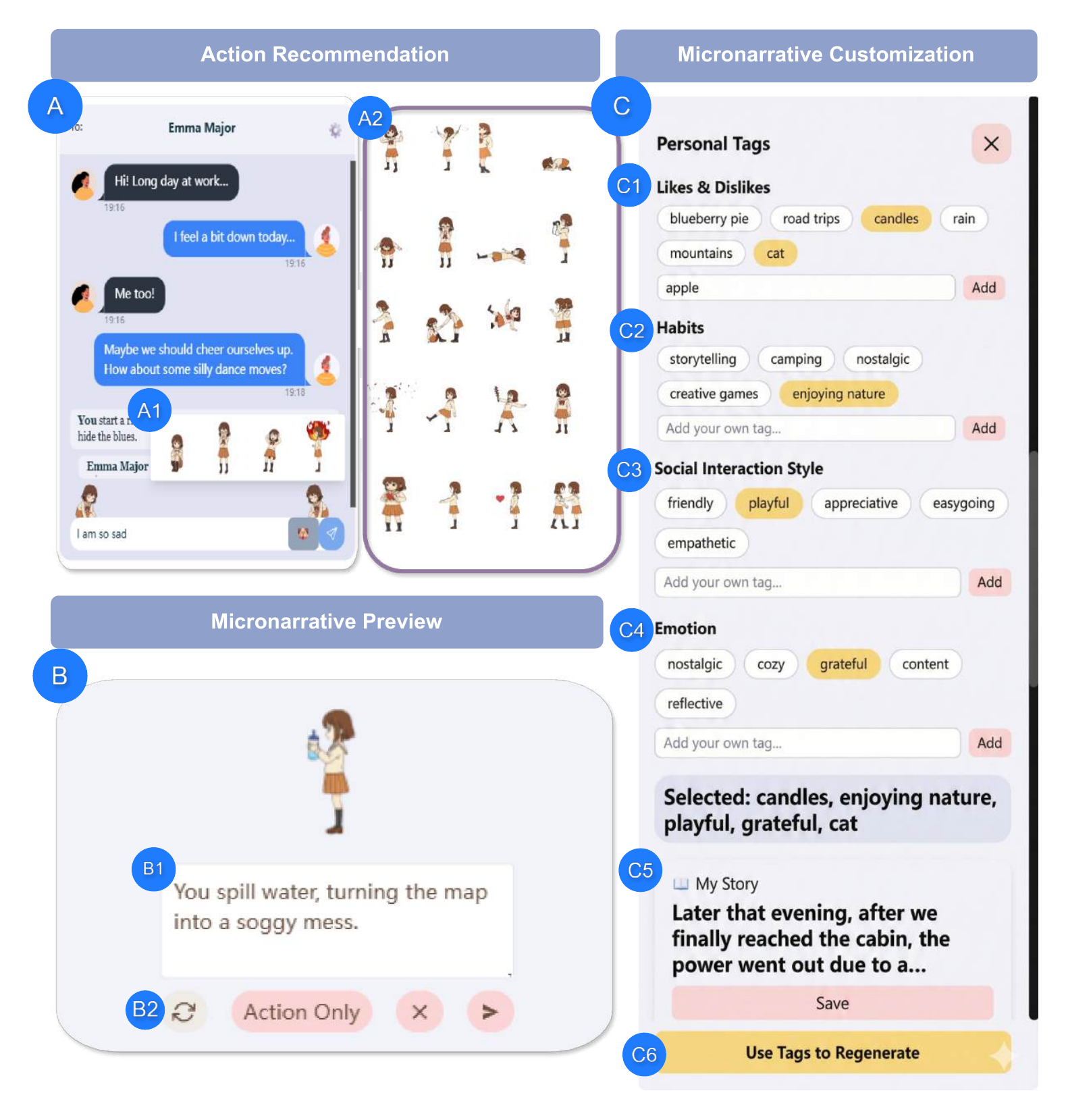}
    \caption{\textbf{Workflow of \emph{PuppetChat}. (A) Everyday chat view. Pressing the Actions button (Fig.~\ref{fig:main_page}~B3) reveals (A1) a recommendation strip with four context aware actions sampled from a 42-item action library (A2), and clicking any action in (A1) advances to (B). (B) Composition view showing (B1) a visual preview of the selected action and an automatically generated micronarrative; tapping (B2) opens the customization panel (C). In (C), users refine tags via (C1) likes and dislikes, (C2) habits, (C3) social interaction style, and (C4) emotion, and may edit (C5) \emph{My Story}. Clicking "Use Tag to Regenerate" (C6) advances to regenerate the micronarrative based on the selected tags and enables sending.}}
    \Description{(A) Everyday chat view. Pressing the Actions button reveals (A1) a recommendation strip with four context-aware actions sampled from a 42-item action library (A2). (B) Composition view showing (B1) a visual preview of the selected action and an automatically generated micronarrative; tapping (B2) opens the customization panel (C). In (C), users refine tags via (C1) likes and dislikes, (C2) habits, (C3) social interaction style, and (C4) emotion, and may edit (C5) \emph{My Story}. Clicking "Use Tag to Regenerate'' (C6) advances to regenerate the micronarrative based on the selected tags and enables sending.}
    \label{fig:workflow}
\end{figure*}

In what follows, we detail \emph{PuppetChat}'s key features, proceeding step by step along the workflow in ~\autoref{fig:workflow}.

\subsubsection{Contextual Action Recommendation (scoring based)}
\label{sec:rec-pipeline}

In our formative study, we found that participants often wanted to trigger puppet actions in two distinct ways: \textbf{with input}, when they typed either explicit commands or affective utterances, and \textbf{without input}, when they simply wished to react in the flow of conversation without sending additional text. Designing for both modes ensures that action recommendations remain responsive across different conversational circumstances.

After the user clicks button B3 in Fig.~\ref{fig:main_page}, the system fuses text understanding and conversational context via a lightweight scoring scheme. For each candidate action $a$ in the action library $\mathcal{A}$ (42 items), it computes a global score
\[
s(a\mid c)
= w_{\text{text}}\,s_{\text{text}}(a\mid c)
+ w_{\text{ctx}}\,s_{\text{ctx}}(a\mid c)
+ w_{\text{pref}}\,\mathrm{Preference}(a,c)
+ \varepsilon(a),
\]
where $c$ is the current context, $s_{\text{text}}(a\mid c)$ is the text based score, $s_{\text{ctx}}(a\mid c)$ is the context based score, $\mathrm{Preference}(a,c)\in[-1,1]$ is a user specific preference term, and $\varepsilon(a)$ is a small exploration noise term. In our implementation, we use fixed weights $w_{\text{text}} = 1.0$, $w_{\text{ctx}} = 1.0$, and $w_{\text{pref}} = 0.5$.

\begin{figure*}[htbp] 
    \centering
    \includegraphics[width=\textwidth]{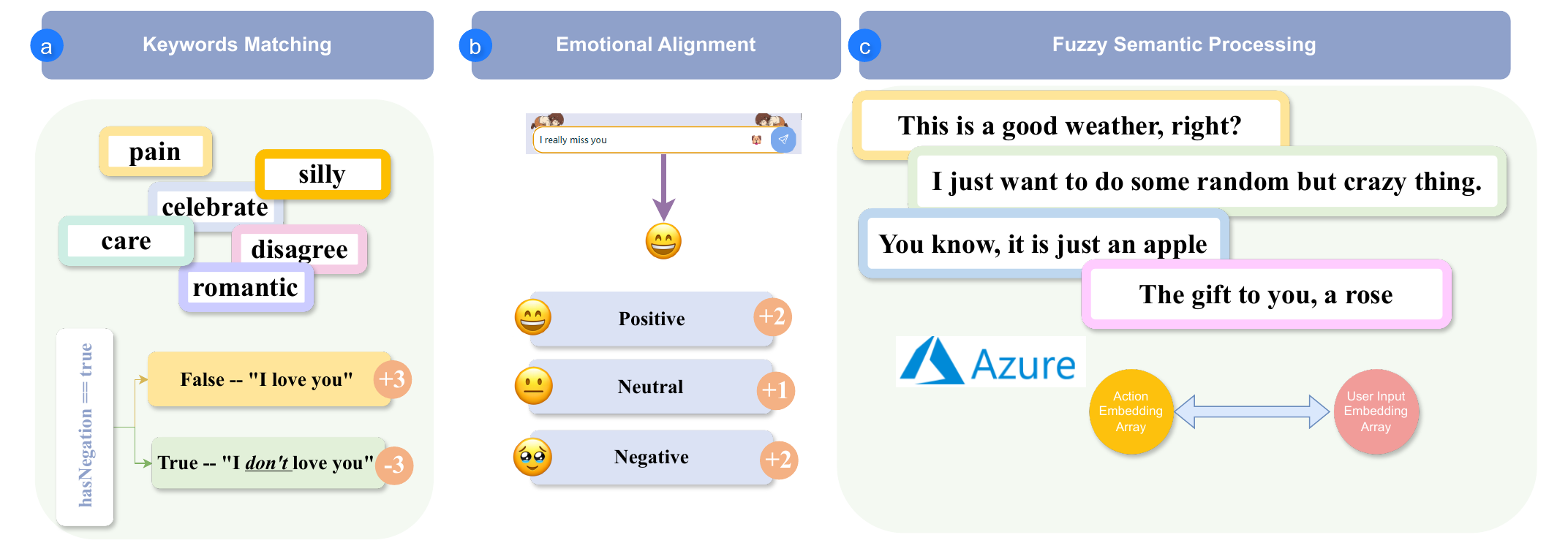}
    \caption{\textbf{Text to action interpretation. (a) Keyword extraction with negation handling aligns user text with action tags; (b) emotional alignment maps input valence (positive, negative, or neutral) to the action's emotion labels; (c) embedding-based fuzzy semantic matching covers indirect or metaphorical expressions.}}
    \Description{Text-to-action interpretation. (a) Keyword extraction with negation handling aligns user text with action tags; (b) emotional alignment maps input valence (positive, negative, or neutral) to the action’s emotion labels; (c) embedding- based fuzzy semantic matching covers indirect or metaphorical expressions.}
    \label{fig:with_input}
\end{figure*}

When the user provides text they wish to send or express, the system performs a three layer interpretation to obtain a text based score $s_{\text{text}}(a\mid c)$ for each candidate action $a$. It first uses a language model to extract keywords and match them to action tags (Fig.~\ref{fig:with_input} a), while performing negation detection to distinguish opposite-polarity statements such as "I love you" (keyword/negation term $+3$; positive valence term $+2$, yielding $s_{\text{text}}\approx +5$ before embedding adjustment) versus "I don't love you" (keyword/negation term $-3$ under negation; negative valence term $+2$, yielding $s_{\text{text}}\approx -1$). It then performs emotional alignment (Fig.~\ref{fig:with_input} b): the model determines the emotional valence of the input (positive, negative, or neutral) and matches it to the action's emotion labels, contributing $+2$ for positive or negative alignment and $+1$ for neutral. Finally, when metaphorical or indirect expressions (e.g., "I will give you an apple") cause the first two steps to default to neutrality, the system computes the cosine similarity between embeddings of the input and the action descriptions/keywords to enable fuzzy semantic matching (Fig.~\ref{fig:with_input} c), adding a small adjustment to $s_{\text{text}}(a\mid c)$.

\begin{figure*}[htbp] 
    \centering
    \includegraphics[width=\textwidth]{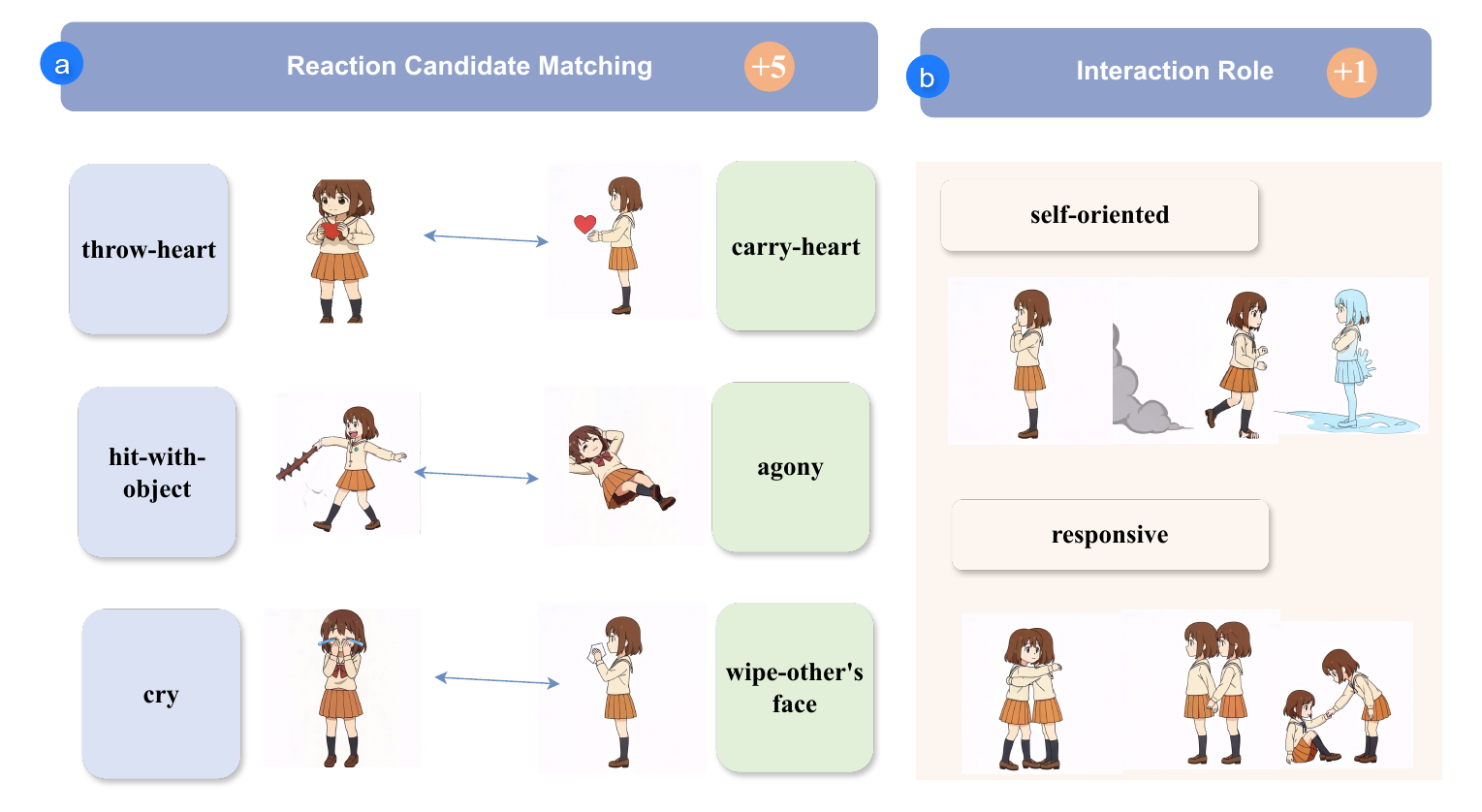}
    \caption{\textbf{Contextual evaluation. (a) Reaction candidate matching: the partner's most recent action queries the action library's \textsc{ReactionCandidates} to prioritize complementary responses; (b) interaction role bias: actions labeled \emph{responsive} (vs.\ \emph{self oriented}) are promoted—conditioned on the ongoing conversational state—to encourage reciprocal exchanges.}}
    \Description{Contextual evaluation. (a) Reaction-candidate matching: the partner’s most recent action queries the action library’s ReactionCandidates to prioritize complementary responses; (b) interaction-role bias: actions labeled responsive are promoted—conditioned on the ongoing conversational state—to encourage reciprocal exchanges.}
    \label{fig:without_input}
\end{figure*}

Users can trigger recommendations in two ways: they may enter text and then press the Actions button (Fig.~\ref{fig:main_page}~B3), or press B3 directly without entering text. In both cases, the contextual channel $s_{\mathrm{ctx}}(a\mid c)$ is evaluated in parallel to ensure responsiveness in reactive situations. As shown in Fig.~\ref{fig:without_input}~a, $s_{\mathrm{ctx}}(a\mid c)$ prioritizes complementary reactions by querying the action library's \textsc{ReactionCandidates} with the partner's most recent action (e.g., \emph{throw heart} $\rightarrow$ \emph{catch/carry heart}, \emph{hit with object} $\rightarrow$ \emph{agony}, \emph{cry} $\rightarrow$ \emph{wipe other's face}); matching candidates receive a +5 bonus. Separately, Fig.~\ref{fig:without_input}~b applies an interaction role bias: each action is labeled as either \emph{self-oriented} or \emph{responsive}, and—conditioned on the ongoing conversational state—those marked \emph{responsive} receive a +1 bonus to encourage reciprocal exchanges. A user specific preference term $\mathrm{Preference}(a,c) \in [-1,1]$ slightly promotes actions the user has frequently selected and demotes those they have repeatedly ignored or hidden. Together with the text based score $s_{\text{text}}(a\mid c)$, these contextual and preference contributions are combined via the scoring function above to yield the final ranking score $s(a\mid c)$ used to select the top four recommendations.

\subsubsection{Micronarrative Generation and Adaptive Regeneration}
\label{sec:micro-regenerate}

While actions enrich expressivity, our formative study showed that users sometimes sought added context or humor to make their actions feel more personal. Prior work on micronarratives suggests that short, situated captions can deepen social presence and support lightweight identity work in digital exchanges \cite{socialpresence02, oh2018systematic}. Building on this insight, \emph{PuppetChat} introduces micronarratives: brief, context aware captions that accompany actions and are generated by drawing on three aspects: 

\begin{enumerate}
    \item the user's personal story, a short self description provided voluntarily (entered in Fig.~\ref{fig:main_page}~B1); 
    \item the selected action, which serves as the expressive trigger; and 
    \item the surrounding conversational context, which situates the action in dialogue. 
\end{enumerate}

When no personal narrative is available, the system defaults to a neutral baseline to preserve coherence. In the workflow, once a user selects any of the four recommendations in Fig.~\ref{fig:workflow}~A1, the composition view in Fig.~\ref{fig:workflow}~B displays the AI-generated micronarrative aligned with that action. Tapping B2 opens the customization panel in Fig.~\ref{fig:workflow}~C, where C1 (likes and dislikes), C2 (habits), C3 (social interaction style), and C4 (emotion) provide lightweight controls for refinement. Conditioned on the personal narrative in Fig.~\ref{fig:main_page}~B1, the system proposes five candidate tags within each category. Users may select from these or add their own. If inspired by the ongoing chat or by chosen tags, users can revise their personal narrative via C5, after which pressing C6 produces a regenerated micronarrative.

Within this process, keywords extracted from the personal narrative act as persona anchors that help maintain a consistent voice across turns while staying sensitive to the current action and context. For example, given the tag "cat" and the action \emph{wipe-tears}, the regenerated caption might read: "I'll be your gentle cat paw wiping away those tears." As another example, with the tag "marathon" and the action \emph{high-five}, the caption could be: "A high-five at mile 26, we made it." These adjustments make captions feel individualized without imposing heavy authoring overhead.

This functionality operationalizes \textbf{DG2} (integrate actions with personalized micronarratives) and \textbf{DG3} (ensure editability and user control over narratives): micronarratives are generated automatically for immediacy yet remain easily rewritable through tags and personal narrative edits. By aligning captions with user provided anchors and conversational context, it also complements \textbf{DG1} by supporting responsive, meaningful exchanges in dyadic chat.

\subsubsection{Reciprocal Action Exchange}
\label{sec:action-exchange}

\begin{figure*}[htbp] 
    \centering
    \includegraphics[width=\textwidth]{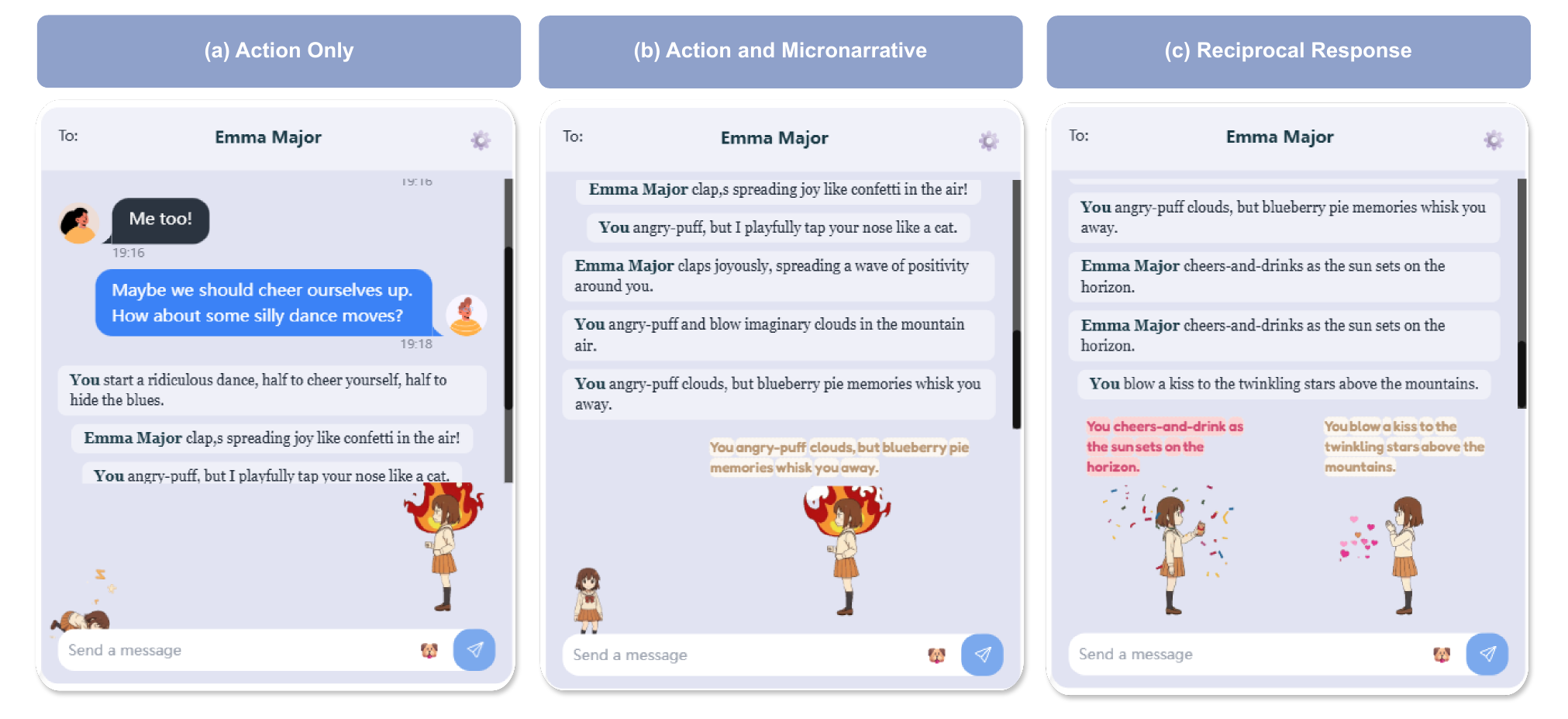}
    \caption{\textbf{Status views for action delivery and exchange. (a) \emph{Action Only}: the selected action is dispatched without a micronarrative. (b) \emph{Send}: the action and the edited micronarrative are delivered together. (c) Partner response: both puppets are co-present and their actions are rendered as an interaction.}}
    \Description{Status views for action delivery and exchange. (a) Action Only: the selected action is dispatched without a micronarrative. (b) Send: the action and the edited micronarrative are delivered together. (c) Partner response: both puppets are co-present and their actions are rendered as an interaction.}
    \label{fig:status}
\end{figure*}

From the workflow controls (Fig.~\ref{fig:workflow} B), the user can either choose "Action Only" or proceed to "Send." Selecting "Action Only" dispatches the chosen action without its micronarrative. The event appears as the ephemeral status view in Fig.~\ref{fig:status} (a) and leaves no micronarrative trace in the log, aligning with DG4 (ephemeral, low commitment expression). Choosing "Send" transmits both the action and the edited micronarrative. The conversation history preserves a persistent textual record as in Fig.~\ref{fig:status} (b). Tapping this record replays the original action rendering.

When the partner responds, the status transitions to a dyadic exchange (Fig.~\ref{fig:status} (c)), in which both puppets are co-present and their actions are rendered as an interaction. The exchange is appended to the thread, supporting turn taking and continuity in the chat history.

\subsection{Implementation}
\subsubsection{Frontend Implementation}
The frontend is built with Vue 3\footnote{https://vuejs.org/} (Composition API) and styled using Tailwind CSS\footnote{https://tailwindcss.com/} for a responsive, utility first interface. It comprises three main components: an action palette, a message input box with micronarrative previews, and a real time activity feed. Client–server communication is maintained via WebSockets\footnote{https://socket.io/} (\texttt{Socket.IO}), ensuring actions and recommendations are broadcast to conversation participants with end-to-end latency of about 200 ms.

\subsubsection{Backend and AI Components}
The backend uses Node.js\footnote{https://nodejs.org/zh-cn} and Express\footnote{https://expressjs.com/}, with REST endpoints for user/session management, CRUD over the action library, and micronarrative generation. Real time events (e.g., \texttt{puppet-action}, \texttt{emn-update}) are broadcast to dyad specific namespaces. Data are stored in MongoDB across collections for actions, conversations, users, logs, and messages. Secondary indexes and attached embeddings (via Azure NLP API\footnote{https://azure.microsoft.com/}) enable semantic retrieval, while Azure OpenAI supports emotion/role inference and micronarrative generation. Recommendations combine heuristics, embedding similarity, and LLM annotations. Results are logged and broadcast, with options for action only or action and micronarrative persistence.

\section{User Study}
We conducted a 10-day field study with 11 dyads (22 participants) in close relationships to assess how \emph{PuppetChat} supports reciprocal action exchanges, personalized micronarratives, and everyday communication\footnote{The study protocol was reviewed and approved by the IRB at the lead author's institution, where the study was conducted.}.

\subsection{Participants}
We recruited 11 dyads (22 participants; 12 women, 10 men; ages 21–28, \emph{M} = 24.45, \emph{SD} = 1.95) in close relationships. Dyads self identified as best friends (6) or couples (5). Participants spanned student and professional roles across HR/business, software engineering, data analytics, banking, nursing, and marketing. To reflect everyday habits across ecosystems, participants reported a heterogeneous mix of messaging applications (e.g., WhatsApp, Instagram, WeChat, Telegram, Discord, LINE, iMessage). We refer to dyads as P1–P11 and to members as A/B; detailed profiles appear in Table~\ref{tab:participants}.

\subsection{Procedure}

Before deployment, each participant completed a baseline questionnaire capturing demographics, commonly used messaging apps, typical communication frequency with a close friend or partner, typical conversation duration, and expected response latency. Participants then used \emph{PuppetChat} in situ with their existing partner or close friend for 10 consecutive days. We provided 10 conversation topics; on Days 1–6, participants selected one topic per day and chatted for at least 15 minutes, while on Days 7–10 they could either continue with listed topics or chat freely without constraints.

During deployment, the system logged 5 fine grained interaction events. In parallel, participants completed a brief two-item micro survey once per day about that day's chatting experience and technical issues. On Day~10, a summative questionnaire assessed \emph{Co-presence and Affective Intimacy} (DG1, DG2), \emph{Reciprocal Interaction} (DG1), \emph{Personalized Micronarratives} (DG2), and \emph{Agency and Persistence} (DG3, DG4), together with the System Usability Scale (SUS) \cite{brooke1996sus, bangor2009determining} and NASA--TLX \cite{hart2006nasa} workload.

Following the deployment, each dyad participated in a 40–60\,min semi-structured interview to probe experiences. 

\subsection{Measures and Data Analysis}

\subsubsection{Quantitative Analysis and Instrument Design} 
\label{sec:questionnare}
Backend telemetry was aggregated into daily and dyad level usage metrics, including action frequency, reciprocal exchange patterns, micronarrative edits, and persistence choices. These data were analyzed using descriptive statistics to capture engagement trends over the 10-day deployment. Standardized usability and workload were assessed using SUS and NASA–TLX scores, computed following standard guidelines.

To capture the specific relational and expressive affordances of the system, we developed a custom 16-item instrument. The design of this instrument followed a deductive scale development approach. We first drew on Social Presence Theory to operationalize co-presence and affective intimacy \cite{kreijns2022social, cui2013building}. For reciprocal interaction, we adapted ideas from turn taking and entrainment in interpersonal coordination \cite{walther1992, walther1996}. Items for personalized micronarratives were informed by research on narrative self disclosure and story based communication in close relationships \cite{digra669}. Finally, our measures of agency and persistence were guided by frameworks in AI Mediated Communication that foreground the balance between algorithmic assistance and user control \cite{walther1996}. Each item was then refined to fit the specific context of PuppetChat and mapped to one of four dimensions: Co-presence and Affective Intimacy (DG1, DG2), Reciprocal Interaction (DG1), Personalized Micronarratives (DG2), and Agency and Persistence (DG3, DG4).

The Basic scale consisted of sixteen items organized into four subdimensions. 
\emph{Co-presence and Affective Intimacy} captured participants' sense of mutual presence and closeness, including perceived partner presence (\emph{Presence}), willingness to disclose genuine feelings (\emph{Disclosure}), the emergence of shared rituals (\emph{Rituals}), and a warmer relational atmosphere (\emph{Warmth}). 
\emph{Reciprocal Interaction} assessed the mutuality of expressive exchanges, including paired or complementary actions (\emph{Pairs}), understanding of action intentions (\emph{Understanding}), appropriateness of recommended reciprocal actions (\emph{Fit}), and the rhythmic quality of matched interactions (\emph{Rhythm}). 
\emph{Personalized Micronarratives} reflected how system generated textual elements enriched continuity, including natural links to past experiences (\emph{Connection}), the construction of coherent stories (\emph{Story}), the possibility of reusing narratives (\emph{Reuse}), and a reverse coded item indicating when narratives felt empty or templated (\emph{Empty}). 
Finally, \emph{Agency and Persistence} captured users' sense of control and temporal durability, including the ease of editing system outputs (\emph{Editing}), feelings of autonomy (\emph{Control}), the ability to preserve meaningful interactions (\emph{Memory}), and a reverse coded item describing ephemeral disappearance of expressions (\emph{Ephemeral}). The final questionnaire items were derived from these same theoretical frameworks, and the complete set of questions is presented in Table~\ref{tab:basic-items}. 

\subsubsection{Qualitative Analysis and Integration} For the qualitative component, we analyzed interviews and open-ended survey responses using a reflexive thematic analysis approach \cite{braun2019reflecting}. Authors independently reviewed all transcripts to generate initial codes related to action exchange, micronarrative use, editing behaviors, friction points, and expectations around persistence. Through iterative discussion, the authors refined these codes, resolving discrepancies and consolidating them into higher level themes. 

\section{Study Results}
We report our study results by combining quantitative snapshots of system use with findings from post study interviews.
\subsection{PuppetChat Usage (N=22)}
Participants' usage logs indicate sustained engagement in composing and refining "action + micronarrative" messages. On average, participants pressed the \emph{Actions} button (Fig.~\ref{fig:main_page} B3) 18.0 times each (\emph{M} = 18.0; median = 17.5; IQR = 13–22.5; range = 8–34) and ultimately sent 14.5 actions (\emph{M} = 14.5; median = 14; IQR = 10.5–18; range = 6–28), yielding an \(\sim\)\textbf{80\%} trigger\(\to\)send conversion. Micronarratives accompanied (Fig.~\ref{fig:status} (c)) 8.2 messages on average (\emph{M} = 8.2; median = 8; IQR = 5–11; range = 2–16), i.e., \(\sim\) 57\% of sent actions. Effort was also evident in editing and "peek behind the scenes" (Fig.~\ref{fig:workflow}~B2): participants carried out \textbf{13.2} edit operations on average (\emph{M} = 13.2; median = 12.5; IQR = 9–16.5; range = 5–27), and partners opened 9.7 previews on average (\emph{M} = 9.7; median = 9.5; IQR = 7–12; range = 3–18). Usage varied widely across individuals—for example, U13 was the heaviest user (28 actions, 16 micronarratives), whereas U20 used the system sparingly (6 actions, 2 micronarratives).

\begin{table}[t]
  \centering
  \small
  \caption{\textbf{PuppetChat usage by participant (original logs).}}
  \Description{PuppetChat usage by participant (original logs).}
  \label{tab:puppetchat-raw}
  \begin{tabular}{lrrrrr}
    \toprule
    Participant & Total & Action & Micronarr. & Edit ops & Preview opens \\
    \midrule
    U01 & 26 & 20 & 12 & 19 & 15 \\
    U02 & 18 & 14 & 9  & 13 & 11 \\
    U03 & 22 & 17 & 10 & 21 & 12 \\
    U04 & 15 & 12 & 5  & 9  & 10 \\
    U05 & 19 & 16 & 11 & 18 & 14 \\
    U06 & 13 & 10 & 6  & 12 & 8  \\
    U07 & 21 & 17 & 9  & 10 & 9  \\
    U08 & 12 & 9  & 4  & 7  & 6  \\
    U09 & 24 & 21 & 13 & 17 & 10 \\
    U10 & 16 & 13 & 7  & 15 & 9  \\
    U11 & 11 & 8  & 3  & 6  & 5  \\
    U12 & 20 & 16 & 8  & 14 & 9  \\
    U13 & 34 & 28 & 16 & 27 & 18 \\
    U14 & 14 & 11 & 6  & 10 & 7  \\
    U15 & 17 & 14 & 8  & 12 & 11 \\
    U16 & 23 & 19 & 12 & 16 & 13 \\
    U17 & 10 & 8  & 5  & 9  & 4  \\
    U18 & 9  & 7  & 4  & 6  & 3  \\
    U19 & 27 & 22 & 14 & 20 & 16 \\
    U20 & 8  & 6  & 2  & 5  & 4  \\
    U21 & 25 & 19 & 11 & 15 & 12 \\
    U22 & 13 & 11 & 5  & 9  & 7  \\
    \bottomrule
  \end{tabular}
\end{table}

\subsection{Quantitative Snapshot from the Basic Scale}
This section reports our results about how participants used editable micronarratives to express subtle emotions and personalize shared meaning.
\subsubsection{Reliability and Descriptive Statistics}
\begin{figure*}[htbp] 
    \centering
    \includegraphics[width=\textwidth]{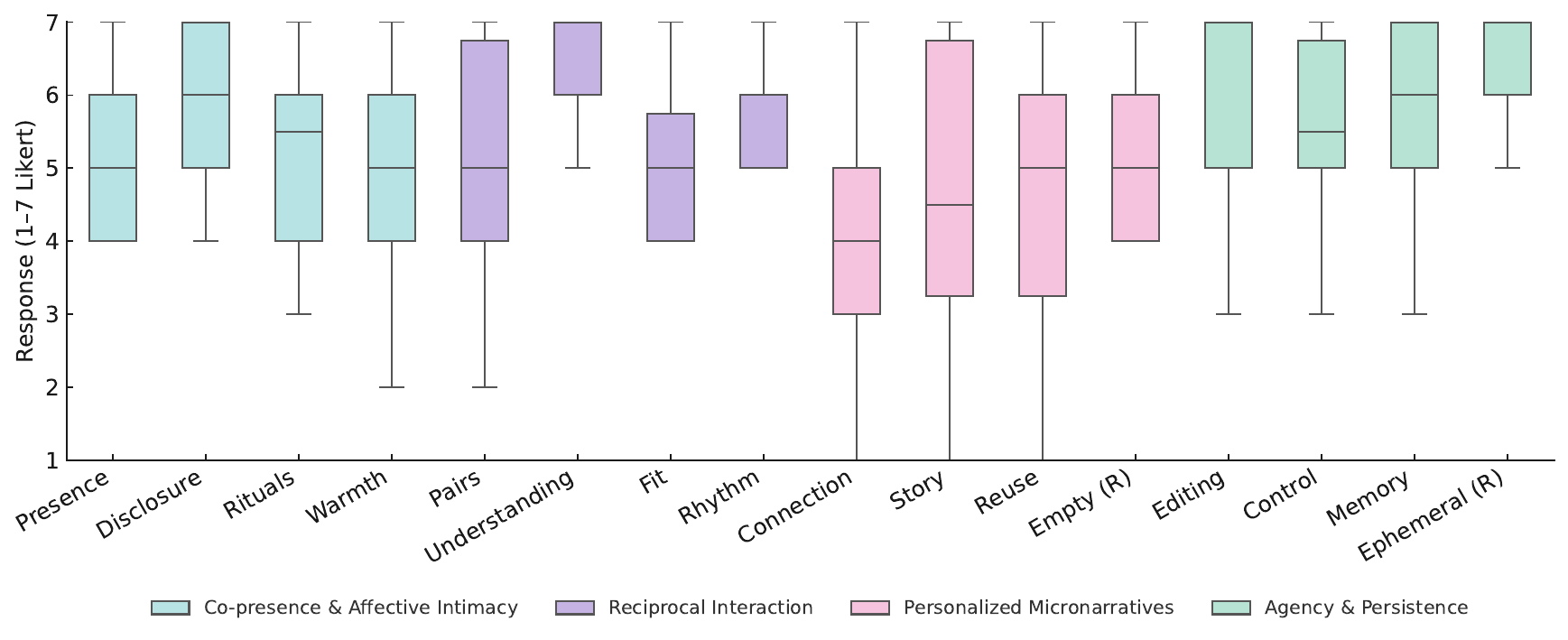}
    \caption{\textbf{Distribution of responses for the 16 items of the \textit{Basic} scale (N=22). 
Each boxplot shows the median and interquartile range of item ratings on a 1–7 Likert scale.  
reverse coded items are labeled with “(R)”; for all other items, higher scores indicate stronger endorsement.}}
    \Description{Distribution of responses for the 16 items of the Basic scale (N=22).
Each boxplot shows the median and interquartile range of item ratings on a 1–7 Likert scale. Reverse-coded items are labeled with “(R)”; for all other items, higher scores indicate stronger endorsement.}
    \label{fig:item-results}
\end{figure*}

We first assessed the internal consistency and descriptive statistics of the four subscales (Table \ref{tab:reliability}) and the overall 16-item instrument(Fig. \ref{fig:item-results}). Cronbach's $\alpha$ values indicated acceptable to excellent reliability across all subscales, with coefficients ranging from .78 to .86, and very high reliability for the full scale ($\alpha = .93$). This suggests that the items within each construct measured coherent underlying dimensions.  

Descriptive analyses showed that participants evaluated the system positively across all dimensions, with mean ratings ranging from 5.1 to 5.4 on a 7-point Likert scale. The overall scale mean was 5.26 ($SD = 1.00$), reflecting generally favorable experiences. Floor effects (scores $\leq 2$) were negligible, while ceiling effects (scores $\geq 6$) were observed in approximately 10--20\% of responses depending on the subscale. These results suggest that while responses were positively skewed, there was sufficient variability to differentiate between participants' experiences.

\begin{table}[t]
  \centering
  \small
  \caption{\textbf{Reliability and descriptive statistics for each subscale and the overall instrument.}}
  \Description{Reliability and descriptive statistics for each subscale and the overall instrument.}
  \label{tab:reliability}
  \begin{tabular}{lrrrrr}
    \toprule
    Scale  & $\alpha$ & $M$ & $SD$ \\
    \midrule
    Co-presence \& Affective Intimacy & .78 & 5.3 & 0.9 \\
    Reciprocal Interaction            & .82 & 5.1 & 1.0  \\
    Personalized Micronarratives      & .86 & 5.2 & 1.1 \\
    Agency \& Persistence             & .83 & 5.4 & 0.9 \\
    \midrule
    Total (16 items)                  & .93 & 5.26 & 1.0\\
    \bottomrule
  \end{tabular}
\end{table}

Figure \ref{fig:item-results} presents a box plot illustrating these results.
Overall, item level means indicated that participants most strongly endorsed 
the persistence related item of resisting ephemerality (\emph{Ephemeral}, $M=6.27$), 
suggesting a clear preference for maintaining interaction records rather than letting them vanish. 
By contrast, the lowest endorsement was observed for the micronarrative item 
linking to past experiences (\emph{Connection}, $M=4.27$), 
highlighting limited resonance between system generated narratives and participants' personal histories. 
Other highly rated items included \emph{Disclosure}, \emph{Understanding}, and \emph{Memory} 
($M > 5.6$), underscoring that PuppetChat effectively fostered openness, reciprocal understanding, 
and the desire to preserve meaningful interactions. 
These results align with the design goals, showing particularly strong support for DG3--DG4 
(Agency and Persistence), while also pointing to opportunities for refining DG2 
(Personalized Micronarratives).

\subsubsection{System Usability and Workload}

To situate our system within established benchmarks, we administered the 
System Usability Scale (SUS; 0--100) and the NASA--TLX workload inventory. 
For the latter, we adopted the raw workload index (RTLX), calculated as the mean 
of the six oriented NASA--TLX subscales (0--100, with the Performance scale reversed). 
Table~\ref{tab:overview} summarizes the results across all participants ($N=22$).

\begin{table}[t]
  \centering
  \caption{\textbf{SUS and NASA–TLX overview. }}
  \Description{SUS and NASA–TLX overview.}
  \label{tab:overview}
  \begin{tabular}{lrrrrr}
    \toprule
    Measure & $N$ & Mean & SD & 95\% CI Low & 95\% CI High \\
    \midrule
    SUS  & 22 & 86.59 & 17.77 & 68.71 & 84.47 \\
    RTLX & 22 & 31.72 & 6.98  & 28.63 & 34.81 \\
    \bottomrule
  \end{tabular}
\end{table}

\begin{figure}[htbp] 
    \centering
    \includegraphics[width=\columnwidth]{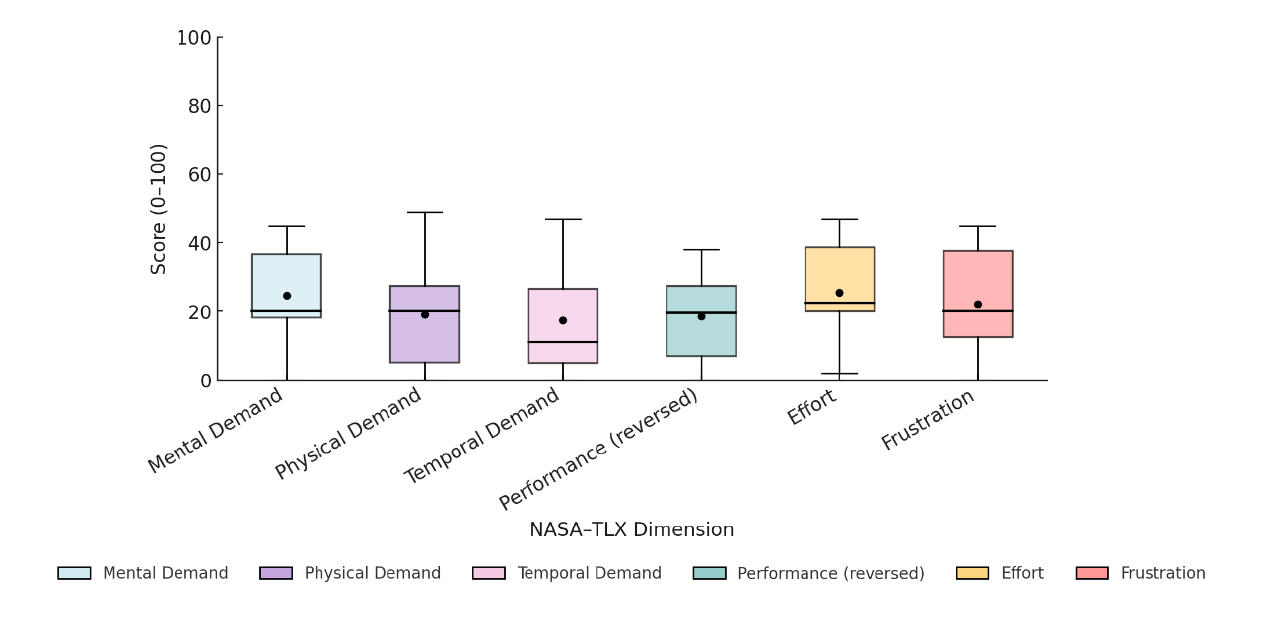}
    \caption{\textbf{Box plots of the six NASA--TLX dimensions.}}
    \Description{Box plots of the six NASA--TLX dimensions.}
    \label{fig:nasa}
\end{figure}

The mean SUS score was $86.59$ ($SD=17.77$), with a 95\% confidence interval of $[68.71, 84.47]$. 
This value substantially exceeds the conventional benchmark of 68, 
and falls into the "excellent usability" range commonly reported for SUS scores above 80. 
In contrast, the mean RTLX workload score (Fig.~\ref{fig:nasa}) was relatively low ($M=31.72$, $SD=6.98$, 95\% CI $[28.63, 34.81]$), 
indicating that interaction with the system required only modest cognitive and physical effort. 

Taken together, these findings suggest that participants experienced PuppetChat as 
both usable and lightweight to operate. The high SUS scores align with the positive 
impressions reported in the qualitative feedback, while the low NASA--TLX scores confirm 
that the system imposed little additional workload beyond the conversational task itself.

\subsection{Post-Study Interviews}
\label{qualitative}
Our post-study interviews with participants further reveal how they relied on PuppetChat to maintain \textit{continuity} over time and reflect on evolving \textit{shared histories}.
\subsubsection{Cultivating Social Presence through Reciprocal Co-performance}

\textit{Establishing social presence hinges on a focused, distraction free communication space and a strong expectation of prompt responses.} Compared with general purpose messengers, PuppetChat's single purpose environment and the "sleeping puppets" cue reduce distraction and heighten joint attention. Participants also reported a strong expectancy that messages would be seen and answered promptly, which stabilized the tempo of interaction and strengthened the felt sense of being co-located. Such as P1-A said \textit{"I really like the two little figures who are always asleep, they make my entry feel meaningful: I can wake them up, which gives me a more focused feeling. Opening PuppyChat feels like stepping into a room that's just ours, no pop-ups, no other chats. "}(P1-A). Another participant reported the same experience, felt a strong sense of confidence and expectation that her partner would quickly notice and respond to her actions. Such as P2-B said \textit{"Because I know that once my action is sent, the other person can respond, and will get many recommended ways to respond. I expect and am confident they'll reply. "} (P2-B)

\textit{Fostering anticipated reciprocity, action response grammar transforms individual expressions into shared, jointly completed performances.} Participants understood the action response grammar as an invitation closure sequence: sending an action initiates a jointly completed performance rather than a solitary broadcast. System supported complementary or mirrored replies make being answered feel probable at the moment of sending, shifting the experience to "we did something together." As P6-A reflected \textit{"Lately I've been exhausted, but I sent a knees-hug action [action: knees-hug] in this app. Getting a hug-back feels like we did something together, not just pasted a sticker. When I sent it, it felt less like expressing my emotion and more like showing my state, and I knew the other person would be there with me."} (P6-A)

\textit{Reanimating shared routines and role complementarity, PuppetChat's co-performance features enable the represencing of familiar life scripts, thereby recalling the embodied sense of us even across distances.} Co-performance is not only emotional, it reanimates shared routines and role complementarity, recalling the embodied "us" despite distance. We found that PuppetChat's co-performance features enabled the represencing of familiar roles and shared routines, creating a strong sense of her partner's presence and recalling past embodied experiences despite physical distance. P10-A said \textit{"Compared to other instant messengers, I feel their presence more strongly here. We hadn't seen each other in person for a long time. I said I had taken many selfies, so I sent a ‘fan-self' action. She replied with ‘take photo,' and I felt like she was still by my side, just like before, my dedicated photographer."} (P10-A)

Cultivating social presence in intimate chat involves establishing a focused and low distraction environment with reliable expectations of reciprocal replies. Upon this foundation, action response turns convert individual expressions into reciprocal, closed sequences of co-performance.  This shift is strongly validated by high mean scores on Co-performance \& Rhythm (M=5.34) and Presence \& Warmth (M=5.24) scales, which also show a strong positive correlation with system usability.

\subsubsection{Expression and Personalization}
\label{expression}
\textit{Editable micronarratives allowed participants to articulate subtle feelings that might otherwise remain unspoken, providing a lightweight way to preserve face while reducing ambiguity.} 
We found that participants used this feature to disclose dissatisfaction or delicate emotions indirectly, articulating what they might hesitate to say outright. P5-A wanted to hint that he was a bit annoyed after his best friend forgot to update him about their weekend plan. Rather than saying this outright, he chose the “agony” action and edited its micronarrative from the dramatic default to a teasing, understated line: “someone went missing today...” This combination let him express himself in a joking and concise way that fit their usual banter. As he explained, \textit{"I'm not someone who speaks very directly, so I edit the short description to express that I'm a little unhappy."} (P5-A)

\textit{Micronarratives also enabled participants to cultivate shared symbolic codes and playful shorthand, reinforcing intimacy and a sense of we-ness.} By pairing actions with personalized text, users generated dyadic "inside jokes" that recalled shared identity and history. We found that such practices generated micro cultures of communication, where reciprocal riffs around everyday dislikes or preferences strengthened relational closeness beyond what emojis or text alone could achieve. As one participant noted, \textit{"We both really hate cilantro, so once I sent a vomiting action, and after she replied with the same action, both of our short descriptions were about cilantro. I was like, wow, this is our tacit understanding, no wonder we're such good friends."} (P9-B)

\textit{Participants diverged in how much effort they invested in editing micronarratives, revealing two contrasting orientations toward use. For some, micronarratives resembled avatars or profile bios, serving as lightweight surfaces to display individuality and highlight shared affinities.} These participants embraced personalization as a way to signal both their personality and their relational commonalities. As P4-A put it, \textit{"I love avatars and profile bios, this micronarrative feels the same. It helps me achieve personalization. On one hand it connects us through common points, but it also lets me show my personality."} (P4-A)
In contrast, others were less inclined to modify text, preferring brevity or shifting emphasis toward the partner rather than themselves. For these participants, too much self related content felt inappropriate or self centered, and they valued contextually relevant prompts that expressed care and attention to the other. As one participant explained, \textit{"I don't really like too much content about my personal information—maybe because I'm the one who fawns over the other person, I often want to express my attention to them. Showing too much of myself can feel self centered. It would be better if it connected more with the conversation context."} (P10-B)

Taken together, these accounts show how PuppetChat supported both personalization oriented identity work and other directed, low effort expression, accommodating different relational and communicative preferences. This resonates with our quantitative results: higher scores on \emph{Editing}, \emph{Control}, and \emph{Memory} (DG3--DG4) confirm strong agency and persistence, while comparatively lower ratings on \emph{Connection} (DG2) highlight the design opportunity to strengthen ties between generated text and users' lived experiences.

\subsubsection{ Continuity and Reflection}
Beyond immediate expression, participants emphasized the importance of continuity and reflection in sustaining meaningful interaction. They viewed PuppetChat not only as a channel for moment-to-moment communication but also as a medium for capturing evolving moods, shared histories, and ongoing narrative arcs. 

\textit{Participants highlighted the need for lightweight tools to maintain continuity, both in tracking daily moods and in preserving a sense of shared history.} Some wished for a "save multiple versions" function for personal stories, treating them as day level state markers that could be easily adjusted without reauthoring from scratch. This versioning function was seen as both expressive, capturing daily shifts in mood, and ergonomic, reducing authoring burden. As one participant explained, \textit{"I wish there were a save button in personal story so I could save my different stories for each day. I treat this description as a reflection of my daily state."} (P3-B). Others emphasized retrospective viewing, using past entries as a jointly owned archive that documents how the relationship evolves over time. As another participant described, \textit{"I revisit past entries because I want to see how my partner and I were in different states at different times—it's a way of preserving memories that belong to us."} (P6-A).

\textit{Continuity also emerged in the form of narrative arcs, where actions and short descriptions clustered around focal themes over stretches of time. }Participants described how prior moves seeded subsequent ones, producing a serialized flow that sustained momentum and deepened engagement with a shared topic. As one participant reflected, \textit{
    "This continuous record feels like a series between us, and I realized our discussion felt like a serialized show. Seeing the descriptions in her replies gave me lots of ideas and made me want to respond."} (P5-A). 

Taken together, these accounts demonstrate how continuity and reflection broadened PuppetChat's role from facilitating single turns of expression to sustaining longer term relational trajectories. The system fostered a sense of temporal depth and shared narrative, reinforcing both personal presence and collective identity over time.

\section{Discussion}
In this section, we interpret our findings, examine how PuppetChat supports reciprocal and expressive communication, outline key design implications, and then reflect on limitations of our study and directions for future work.

\subsection{From Terminal Feedback to Reciprocal Co-Performance}
\label{sec:terminal-feedback}
A central finding of our work is that intimacy in mediated environments is structurally supported by \textit{invitational affordances}, interface mechanisms that solicit and structure reciprocal responses rather than by mere visual richness. Prior literature on IM often highlights the prevalence of "terminal feedback", such as reaction emojis that signal acknowledgement but effectively close a thread \cite{coherence01, schwarz2011moved}. In contrast, PuppetChat's complementary action recommendation system (Fig.~5 (a)) offers a concrete instantiation of a \textit{grammar of interdependence}: suggested responses are framed as follow-ups that fit the partner's move rather than as isolated reactions. The quantitative results, particularly the high scores on \textit{Reciprocal Interaction} ($M=5.1$) and \textit{Co-presence} ($M=5.3$), are consistent with the interpretation that users perceived these exchanges not as isolated signals, but as coupled units of behavior.

We interpret this shift as the emergence of \textit{asynchronous co-performance}: a form of joint action in which temporally staggered, yet mutually responsive, turns accumulate into a shared performance. Qualitative accounts of the "sleeping puppets" creating a "focused room" suggest that the interface staged a \textit{dormant co-presence}—a state of readiness that elevated the expectation of a reply even when partners were not strictly co-temporal. Unlike standard stickers, where the choice is vast and often arbitrary, the structured recommendations narrowed the design space to actions that "fit" the partner's previous move (e.g., distress $\rightarrow$ comfort). This constraint appeared to increase the perceived meaningfulness of the interaction: by reducing the cognitive load of selecting a response, the system allowed users to focus on the \textit{rhythm} of the exchange. This reading aligns with theories of joint action \cite{jointaction01}, suggesting that in remote contexts, the feeling of "being together" can arise less from strict simultaneity and more from the \textit{complementarity} of contributions.

The same invitational structure appears to underpin continuity. Rather than terminating an exchange, reciprocal chains of actions and replies keep threads alive across moments, while editable micronarratives let users quickly tune meaning without reauthoring from scratch. When users choose to persist action and micronarrative pairs, these micro sequences stitch across sessions into lightweight, revisitable traces. This pattern is consistent with prior work on how persistent traces and joint reminiscence scaffold shared memory and social presence \cite{alea2007ll, socialpresence02}, showing that selected exchanges become a jointly owned archive that can be replayed to re-experience prior moods and rhythms and to pick up ongoing threads with a sense of being seen across time.

Our deployment included both romantic partners and best friends, all of whom described their relationships as already close. We did not design or power the study to test systematic differences between relationship types, and we observed no robust quantitative divergences in how reciprocal actions or continuity features were taken up between close friends and romantic partners. Any descriptive differences (e.g., in thematic content or humour style) were minor. We therefore cautiously treat these mechanisms as applying to close dyads more generally, while acknowledging that relationship type, history, and stage of intimacy may modulate how strongly specific features (e.g., playful co-performance vs.\ reflective revisiting of traces) are used. A finer grained comparison of early stage versus established relationships, and of different relational forms, remains an important direction for future work rather than a claim of this study.

\subsection{Micronarratives: Ambiguity and Specificity}
While embodied actions provided the vehicle for co-presence, micronarratives—short, AI suggested captions paired with those actions provided the semantic anchor. Our results indicate a complex relationship between AI generated text and user intent. The "cilantro" example (Section~\ref{expression}), in which a generic template about food was appropriated into a highly specific inside joke, illustrates how dyads repurposed generic suggestions to construct private meaning. This supports the notion that intimacy tools should leave room for multiple interpretations \cite{baxter1986turning}, but adds a nuance: AI generation in this setting serves as a \textit{scaffold for appropriation} rather than a final product.

A potential concern is that AI generated micronarratives might feel impersonal or disconnected from the relationship, or that reliance on such suggestions could erode people's ability to craft expressive conversations on their own. The relatively modest score on the \textit{Connection} item, which captured how well the system linked to past experiences, reflects this limitation and suggests that generic templates alone may not fully convey relational history. At the same time, participants frequently engaged with the editing tools and reported high willingness to disclose, suggesting that they did not treat the initial AI output as a finished message. Instead, “imperfect” suggestions functioned as provocative resources that invited users to articulate and refine feelings they might otherwise leave unspoken.

Here, the value of micronarratives lies less in the accuracy of zero shot generation and more in lowering the activation energy for nuanced, face saving self disclosure, making it easier to start and adjust a vulnerable message without feeling overly exposed. Although our 10 day deployment cannot address long term skill change or subtle shifts in authorship, the observed pattern points toward \textit{scaffolding} rather than \textit{substitution}: meaningful exchanges emerged from users' active reshaping of generated text, indicating that partners continued to practice and elaborate their own relational language rather than outsourcing it to generative AI. Longer term, comparative studies are needed to examine how such scaffolds affect people's expressive repertoires over time.

\subsection{Design Implications}
The design and evaluation of PuppetChat provide several design implications for HCI and CMC systems seeking to support remote intimacy: 
\begin{itemize}
\item Embed embodied actions into messaging turn structure with complementary grammars. Rather than treating gestures, animations, or haptics as one-off reactions, interfaces can define simple complementary patterns (e.g., distress $\rightarrow$ comfort, offer $\rightarrow$ acceptance) that make a reply feel both easy and expected, turning individual expressions into co-performances. Our participants' descriptions of "sleeping puppets" and back-and-forth exchanges suggest that such patterns can support a felt sense of asynchronous togetherness.
\item Offer a lightweight, editable middle layer of expression. Short, regenerable captions paired with actions can bridge the gap between emojis and full text, especially when they support face saving specificity and inside jokes. This layer should be easily skippable for users who prefer minimalism, but powerful for those who wish to craft a distinctive persona and shared humour, as reflected in how dyads appropriated and edited micronarratives in our study.
\item Prioritize user agency in AI mediation. AI should propose rather than impose: recommendations must be editable, overridable, and tunable in intensity, with clear affordances for decline. Grounding generation in user supplied tags and stories respects privacy while still enabling personalization. Our observations of frequent editing and selective uptake highlight the importance of keeping users in control of how AI generated text enters their relational language.
\item Design persistence as narrative infrastructure. Giving users nuanced control over which interactions persist and how they can be replayed, grouped, or revisited can help transform momentary exchanges into a shared history. Simple mechanisms such as replay on click and optional story versioning can support reflection and joint reminiscence without heavy authoring, echoing our participants' use of persisted action+micronarrative pairs as lightweight, revisitable traces.
\end{itemize}
Although our deployment involved a mix of romantic partners and close friends, the underlying mechanisms proved meaningful across relationship types as discussed in section \ref{sec:terminal-feedback}. Rather than being tied to a specific relational category, these implications point to generalizable interaction principles for supporting care, play, and a sense of co presence at a distance.

\subsection{Limitations and Future Work}

While our field study deployment demonstrates the potential of \emph{PuppetChat} in enhancing intimate communication among already close dyads, several limitations remain. First, the relatively small sample size of 22 participants, recruited primarily from a university community and consisting of long term friends and romantic partners, constrains the generalizability of our findings. We did not systematically compare different relationship forms or stages of intimacy, nor did we explicitly recruit neurodivergent participants or attend to intercultural variation in depth. Second, the deployment spanned only ten days, so we cannot speak to the long term impact of reciprocal actions and story driven micronarratives on relationship trajectories, nor can we assess whether people might come to over rely on micronarratives in ways that affect their in person communication. Finally, the study evaluated the integrated system as a whole without comparison conditions, making it difficult to identify which specific elements (e.g., \textsc{ReactionCandidate}, micronarrative generation, editing tools, persistence settings) contributed most strongly to the observed effects.

Future work will therefore expand both scope and granularity. We plan to broaden the participant pool to include a more diverse range of partnerships (e.g., family members, intercultural and cross border relationships), and to examine how different relationship types and intimacy stages modulate the uptake and impact of co-performance and continuity mechanisms. Although our present study does not speak directly to these populations, we see particular promise in studying how action and micronarrative scaffolds can support neurodivergent individuals who may find it challenging to communicate or interpret tone and intent in text based messaging, while carefully attending to accessibility and agency. Cross cultural deployments will be important for understanding how norms of expressivity, face saving, and reciprocity shape the design space. Methodologically, we aim to conduct longitudinal studies to assess \emph{PuppetChat}'s longer term influence on relational dynamics, and to perform ablation and comparison studies (e.g., disabling complementary recommendations, removing micronarratives, or varying persistence) to clarify which mechanisms, individually and in combination, are most impactful for future system design.

\section{Conclusion}
In this paper, we present PuppetChat, a dyadic messaging prototype that enriches intimate communication by addressing the limits of one-off reactions. Users express with lightweight character actions coupled to editable, context aware micronarratives, while the system embeds structured reciprocity by surfacing complementary or mirrored responses at send time, turning solitary expressions into reciprocal sequences. Selective record and replay then supports continuity and shared memory. Our user study demonstrated that PuppetChat supports users in initiating richer conversations and adapting AI suggestions to their own relational tone, as well as maintaining a sustained sense of connection over time, resulting in more engaged and meaningful exchanges compared to everyday messaging practices. 

\bibliographystyle{ACM-Reference-Format}
\bibliography{sample-base}

\appendix

\section{Questionnaire}
To assess participants’ subjective experiences with the system, we developed a questionnaire based on four conceptual subscales reflecting key interaction qualities. Each subscale consists of multiple facets capturing distinct aspects of the experience, with both positively and negatively worded items included. The full set of items is summarized in \textbf{Table~\ref{tab:basic-items}}.

\begin{table*}[t]
\centering
\caption{\textbf{BASIC scale items by subscale and facet. (R) indicates reverse coded items.}}
\Description{BASIC scale items by subscale and facet. (R) indicates reverse-coded items.}
\label{tab:basic-items}

\small
\begin{tabular}{p{4.0cm} p{1.5cm} p{10.5cm}}
\toprule
Subscale & Facet & Item text \\
\midrule
\multirow{4}{*}{Co-presence and Affective Intimacy} 
  & Presence & When using PuppetChat, I often feel as if my partner is right in front of me. \\
  & Disclosure & In conversations, I am more willing to share my genuine feelings or updates. \\
  & Rituals & We developed some expressions or rituals unique to our relationship. \\
  & Warmth & Compared to our usual chatting style, our relationship feels warmer. \\
\midrule
\multirow{4}{*}{Reciprocal Interaction} 
  & Pairs & We often exchange paired or complementary actions. \\
  & Understanding & I can easily understand the intention behind my partner's actions and respond accordingly. \\
  & Fit & The reciprocal actions recommended by the system are usually appropriate and easy to use. \\
  & Rhythm & The combinations and matches of interactive actions make our conversations more rhythmic and enjoyable. \\
\midrule
\multirow{4}{*}{Personalized Micronarratives} 
  & Connection & The system generated micronarratives often naturally connect to my past experiences. \\
  & Story & These micronarratives help us link fragmented conversations into more continuous little stories. \\
  & Reuse & I would like to reuse previous micronarratives as a form of continuity. \\
  & Empty (R) & Some micronarratives feel empty, like templates. \\
\midrule
\multirow{4}{*}{Agency and Persistence} 
  & Editing & I can easily edit or rewrite the actions or micronarratives recommended by the system. \\
  & Control & This editing ability makes me feel more comfortable and in control. \\
  & Memory & For meaningful or memorable interactions, I prefer that they can be saved and revisited later. \\
  & Ephemeral (R) & I prefer that all expressions quickly disappear without leaving any record. \\
\bottomrule
\end{tabular}
\end{table*}

\section{Evaluation}
We conducted a user study with 11 dyads to evaluate the system in realistic interpersonal communication contexts. Participants varied in age, relationship type, occupation, and primary messaging platforms, providing a diverse sample of everyday messaging practices. Detailed participant demographics and background information are shown in \textbf{Table~\ref{tab:participants}}.

\begin{table*}[t]
\centering
\caption{\textbf{Participants (11 dyads): demographics, relationship, occupation, and messaging apps}}
\Description{Participants (11 dyads): demographics, relationship, occupation, and messaging apps}
\label{tab:participants}
\small
\begin{tabular}{l l c l l l}
\toprule
ID & Gender & Age & Relationship & Occupation & Messaging Apps \\
\midrule
P1-A & Female & 24 & Best friends & Student & WeChat; WhatsApp; Instagram \\
P1-B & Female & 23 & Best friends & Student \\
P2-A & Female & 25 & Couple & HR Business Partner & WeChat; WhatsApp \\
P2-B & Male   & 26 & Couple & Software Developer    \\
P3-A & Female & 25 & Best friends & Student & LINE; Instagram \\
P3-B & Female & 22 & Best friends & Student  \\
P4-A & Female & 24 & Couple & Data Analyst        & iMessage; WhatsApp \\
P4-B & Male   & 26 & Couple & Bank Staff          \\
P5-A & Male   & 23 & Best friends & Student & Discord; Telegram \\
P5-B & Male   & 23 & Best friends & Student  \\
P6-A & Female & 27 & Couple & Nurse               & WhatsApp; Signal \\
P6-B & Male   & 27 & Couple & Civil Engineer       \\
P7-A & Female & 21 & Best friends & student & Snapchat; Instagram \\
P7-B & Female & 21 & Best friends & student \\
P8-A & Male   & 28 & Couple & Software Engineer     & WeChat; Telegram \\
P8-B & Female & 27 & Couple & Software Engineer          \\
P9-A & Male   & 26 & Best friends & Software Engineer & Instagram; Signal; Twitter/X \\
P9-B & Male   & 25 & Best friends & Data Scientist     \\
P10-A & Female & 23 & Couple & Research Assistant  & WhatsApp; Facebook Messenger \\
P10-B & Male   & 24 & Couple & Student          \\
P11-A & Male   & 24 & Best friends & High School Teacher & LINE; Telegram \\
P11-B & Female & 24 & Best friends & Marketing Specialist \\
\bottomrule
\end{tabular}
\end{table*}

\end{document}